\documentclass[useAMS]{gGAF2e}

\begin{document}

\newcommand{\Ru}{\mathrm{Re}}
\newcommand{\Rm}{\mathrm{Rm}}
\newcommand{\Pm}{\mathrm{Pm}}

\doi{10.1080/03091920xxxxxxxxx} \issn{1029-0419} \issnp{0309-1929}
 \jvol{00} \jnum{00} \jyear{2006} \jmonth{February}

\markboth{VLADIMIR BATALOV, ANDREY SUKHANOVSKY and PETER FRICK
}{Laboratory study of differential rotation in a convective
rotating layer}

\title{Laboratory study of differential rotation in a convective
rotating layer}
\author{VLADIMIR BATALOV, ANDREY SUKHANOVSKY $\dagger$ and
PETER FRICK \\ Institute of Continuous Media Mechanics, Korolyov
Street 1, Perm, 614013, Russia \thanks{ \vspace{6pt}
\newline{\tiny{ {\em }$\dagger$ Corresponding author: san@icmm.ru }}}}

\maketitle

\begin{abstract}
The evolution of a large-scale azimuthal velocity field in a
rotating cylindrical layer of fluid (radius 150 mm, depth 30 mm,
free upper surface) with meridional convective circulation was
studied experimentally. Two cases were considered: direct
circulation provided by a rim heater at the periphery and indirect
circulation provided by a central heater. The heating rate is
characterized by the Grasshoff number ${\rm Gr_f}$ defined through
the density of the heat flux. The detailed 3D structure of the
mean large-scale velocity field is reconstructed using the PIV
technique for $10^5< {\rm Gr_f}<4\cdot 10^7$. It was shown that
the energy of meridional circulation grows with the Grasshoff
number as $\sqrt{\rm Gr_f}$ in both directions of circulation. Due
to the action of the Coriolis force the meridional flow results in
differential rotation. Differential rotation is characterized by
the mean values of radial and vertical gradients of azimuthal
velocity. Strong negative mean radial gradient which grows with
the Grasshoff number is provided by direct circulation. In the
case of indirect circulation a pronounced negative gradient arises
at moderate Grashoff number. The behavior of the mean vertical
gradient is quite different: a positive vertical gradient grows
logarithmically with Grasshoff number under direct circulation,
whereas a weak negative gradient characterizes the indirect
circulation. This difference follows from the structure of the
flow -- the direct circulation provides a large cyclonic area
localized above the anticyclonic flow, while indirect circulation
leads to a strong separation of these two areas in the radial
direction (the central part is occupied by the cyclonic flow and
the periphery by the anticyclonic flow). Meridional circulation
leads to substantial variation of the integral angular momentum.
Direct circulation results in the growth of the integral angular
momentum and indirect circulation causes it to decrease. At the
same heating power, the increase of angular momentum at direct
circulation is much stronger than its decrease at indirect
circulation.
\end{abstract}

\section{Introduction}

All cosmic bodies rotate to a greater or lesser extent. The
evolution of these bodies (excluding catastrophic events) proceeds
under a very slow variation of the global angular momentum.
However, if the cosmic object is gaseous or liquid (galaxies,
stars) or includes liquid and/or gaseous shells (planets) the
averaged large-scale azimuthal velocity field becomes essentially
inhomogeneous in both the radial and meridional directions. The
deviation from the solid body rotation is called differential
rotation (DR).

DR plays a crucial role in the generation of cosmic magnetic
fields \citep{zrs83} and has created considerable interest in
investigating DR in electrically conducting cosmic media - such as
stellar convective zones, liquid planet cores and galactic disks
\citep{kleeorin, kitchatinov_rudiger}. The study of DR in
nonconducting rotating (spherical) layers has primarily been
motivated by atmospheric research because DR is a part of the
global atmospheric circulation that determines climate formation
\citep{williams2}. For the most part, large-scale circulation in
the atmosphere is caused by the horizontal temperature gradient,
i.e., it has a convective nature. This strongly motivated
laboratory experiments on convection in rotating vessels
(cylinders or annuli) \citep{Killworth, Golitsyn}. The first
generation of laboratory experiments primarily intended  to
reproduce the Hadley cell (or direct meridional circulation) by
realizing a convective flow in rotating annuli or flat cylinders
heated at the periphery and cooled in the center \citep{Hide,
fultz1}. Wave-transition spectra and vacillations were studied by
\cite{fultz2}. Interest in DR in this configuration has been
heightened by the discovery of so-called super-rotation in the
atmosphere of Venus \citep{belton} - the existence in the high
atmosphere of very fast latitudinal circulation, which exceeds by
about 60 times the rotational velocity of the surface (the
rotation period is about four days versus 243 days of the Venus
astronomical day). The problem of super-rotation is directly
linked with processes of angular momentum diffusion and transport
in planetary atmospheres. It was proposed by \cite{gierash} that
Venus super-rotation is maintained by angular momentum transport
provided by meridional circulation. Some mechanisms, which may
contribute to the development of a strong differential rotation
were proposed \citep{Shubert, Rossow}. Comprehensive review of
different quasy-axisimmetric models of planetary atmospheres can
be found in \citep{Read1}.

The first experimental attempt to quantitatively analyze
meridional transport of angular momentum in the rotating layer of
fluid in the case of direct meridional circulation was realized by
\cite{fultz2}, however, integral characteristics such as overall
angular momentum were not studied. Numerical simulations for a
similarly stated problem using different boundary conditions were
presented by \cite{williams1, williams2} and showed good agreement
with experimental results from \cite{fultz2}.

Properties of angular momentum budget and super-rotation of
axisymmetric thermally-driven circulations in a rotating
cylidrical fluid annulus were studied numerically by \cite{Read2}.
Angular momentum diffusion due to molecular viscosity was examined
and was found to be very important for super-rotating flow
formation in a system with stress-free top and side boundaries and
a non-slip bottom. It was shown that global super-rotation is
connected with direct meridional circulation.

 An opposite situation to the direct meridional cell is
the indirect cell, usually produced in a rotating layer of fluid
by heating at the center. Mainly, it was studied in the context of
another atmospheric problem concerning the generation of intense
large-scale vortices. Laboratory modeling of these typhoon-like
vortices was made \citep{Bogatyrev,BogatyrevSmorodin,
BogatyrevLevina et al}. These studies were concentrated on
cyclonic vortex formation and its evolution. A qualitative
experimental study of convective flow driven by a finite-sized
circular heating plate at the bottom of a horizontal fluid layer,
both with and without background rotation, was carried out by
\cite{Boubnov}. The angular momentum budget for an indirect cell
was analyzed by \cite{Read2} and it was shown that indirect
circulation resulted in local and/or global sub-rotation, thereby
demonstrating that the change in overall angular momentum
essentially depends on the type of meridional circulation.

In this paper, we return to experimental investigation of the
convection flow in rotating cylindrical vessels, being motivated
by following reasons. First, we examine the differential rotation
problem in the rotating layer heated at the periphery or center
from a uniform viewpoint, paying special attention to the integral
characteristics of differential rotation. Second, our experiment
concerns the {\it shallow} layer of fluid. And third, we obtain
the full structure of meridional and azimuthal velocity fields
exploiting the potentialities of the PIV technique \citep{PIV,
Ke-Qing Xia}.

We note that our statement of the problem is similar to that made
in \citep{Read2}, except for the aspect ratio and the physical
properties of the fluid. Numerical simulations in \citep{Read2}
were carried out for fluid layers in an annular channel with an
aspect ratio close to unity, whereas we studied the cylinder layer
(no inner wall) with a fixed aspect ratio ${\varepsilon} = h/R =
0.2$ ($h$ - depth of the layer, $R$ - radius of the cylindrical
vessel). The Prandtl number for the fluid in \citep{Read2} was
about 10, and in our study it is about 100.

The structure of the paper is as follows. The experimental set up
is described in Sec.~\ref{setup}. The structure of convective
flows for direct and indirect meridional cells is represented in
Sec.~\ref{dir_circ} and Sec.~\ref{inv_circ}, respectively.
Integral characteristics of DR are described in
Sec.~\ref{integral}. Our results are summarized in
Sec.~\ref{diss}.

\section{Experimental set-up}
\label{setup}

\begin{figure}
\begin{center}
\includegraphics[width=.45\textwidth]{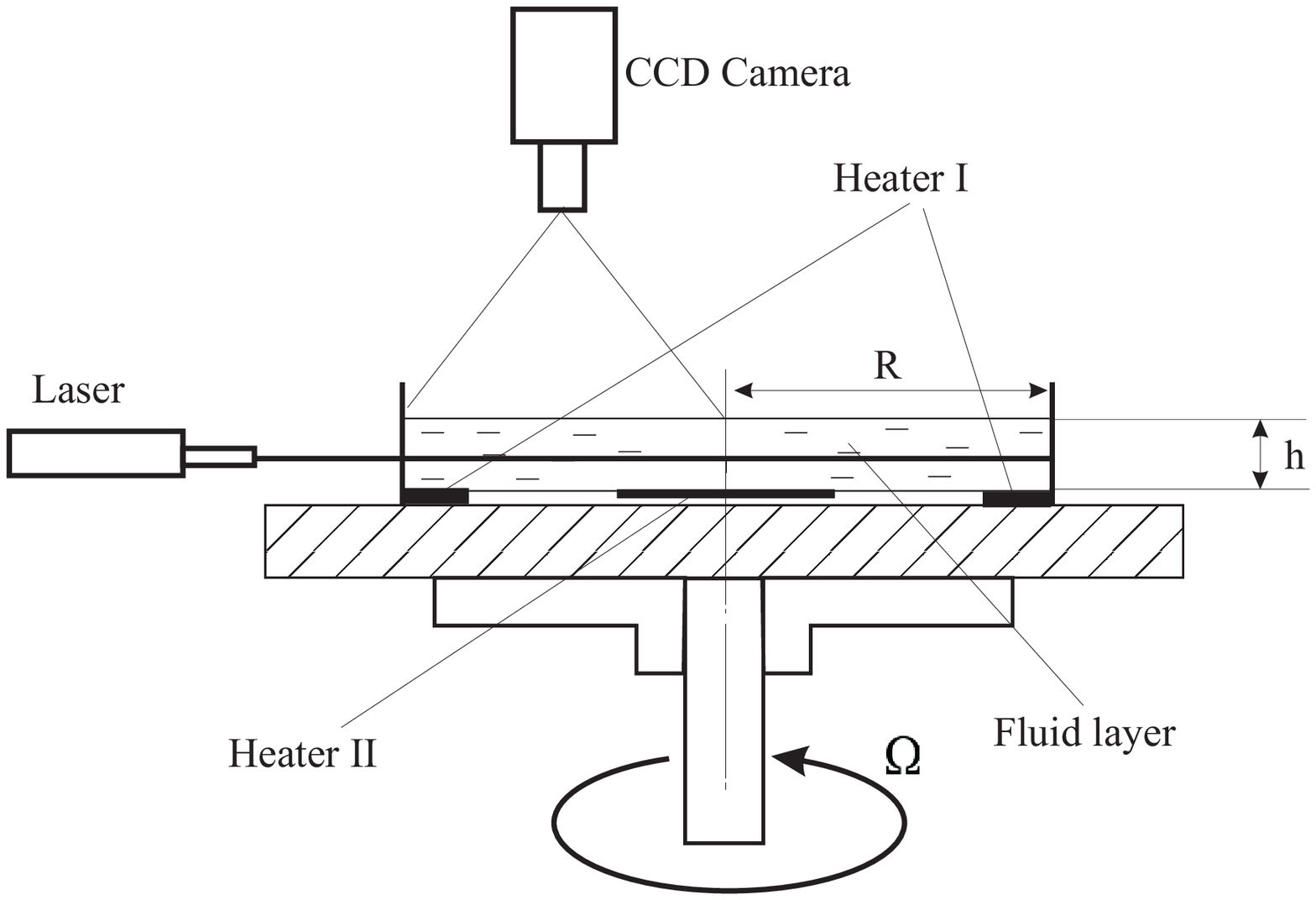}
\includegraphics[width=.25\textwidth]{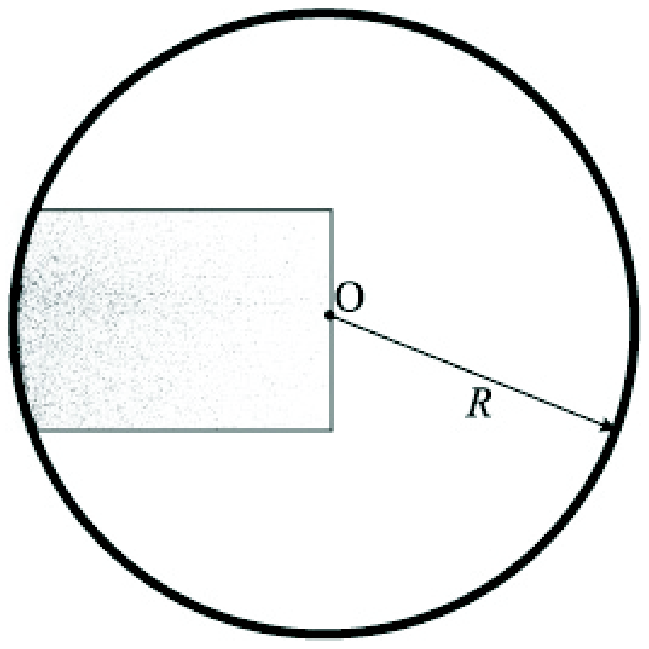}
\caption{Left panel - scheme of the  laboratory set-up; Right
panel (view from above) - zone of horizontal velocity measurements
where a typical (here inversed) PIV image is shown. O - axis of
rotation.} \label{fig_Setup_Field}
\end{center}
\end{figure}

We studied a convective flow in a flat cylindrical vessel placed
on a rotating horizontal table(Fig.\,\ref{fig_Setup_Field}). The
table provides uniform rotation in the angular speed range
$0.04\leq\Omega\leq 0.30$ s$^{-1}$ (with accuracy of $\pm0.002$
s$^{-1}$). The cylinder radius is $R=150$ mm. The cylinder is made
of acrylic plastic, which is transparent to the laser sheet. The
fluid is transformer oil, characterized by a high Prandtl number
\begin{equation}
{\rm Pr} = {\nu}/{\chi}, \label{Prandtl}
\end{equation}
where $\nu, \chi$ are the kinematic viscosity and thermal
conductivity (for transformer oil, ${\rm Pr}=116$ at 50$^{0}$ C).
The layer thickness is constant in all experiments ($h=30$ mm) and
the upper surface is free.

\begin{figure}
\begin{center}
\includegraphics[width=180mm]{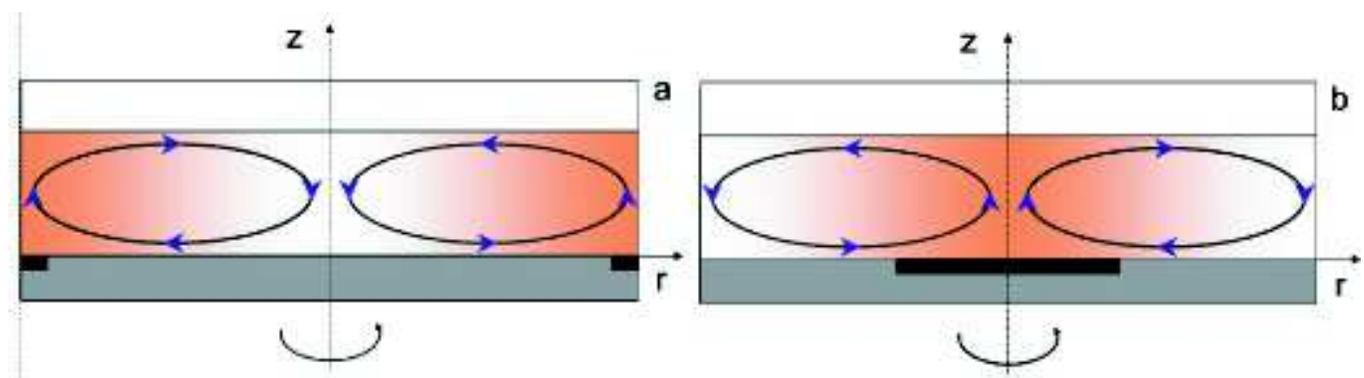}
\caption{Sketch of meridional circulation, initiated by peripheral
heater I (direct circulation - a) and  by central heater II
(indirect circulation - b). The heaters are shown by black boxes.
} \label{circul}
\end{center}
\end{figure}

Two heaters were installed at the bottom of the vessel
(Fig.\,\ref{fig_Setup_Field}). Heater I is a copper rim 20 mm wide
placed at the periphery of the cylinder. This heater provided the
direct meridional circulation as shown in Fig.\,\ref{circul}(a).
Heater II is a copper cylinder of radius 52 mm, installed in the
center of the vessel and coinciding with the axes of rotation. The
heating near the rotation axes (polar heating) provides
circulation in the opposite direction (indirect circulation,
Fig.\,\ref{circul}(b)). Both heaters operate from a DC power
supply. The room temperature is kept constant, and cooling is
provided by the heat exchange on the free surface. It takes about
two hours to obtain a steady-state temperature regime.

\begin{table}
\begin{center}

\caption {Experimental parameters} \label{tab:Exp}

\begin{tabular}{|c|c|c|c|c|c|} \hline
 Heater & $P, Wt$  & $\Omega$ & $E$ & ${\rm Gr_f}$ & ${\rm Pr}$ \\ \hline
 I & 14.2  & 0.069 & 0.118 & $1.4\cdot10^{5}$ & 209\\ \hline
 I &  35.3  & 0.069 & 0.082 & $7.7\cdot10^{5}$ & 151\\ \hline
 I &  55.1  & 0.069 & 0.053 & $2.9\cdot10^{6}$ & 101\\ \hline
 I &  73.8  & 0.069 & 0.044 & $5.8\cdot10^{6}$ & 85\\ \hline
 I &  95.8  & 0.069 & 0.032 & $1.4\cdot10^{7}$ & 63\\ \hline
 II &  6.9  & 0.069 & 0.137 & $1.1\cdot10^{5}$ & 226\\ \hline
 II &  14.3  & 0.069 & 0.104 & $4.1\cdot10^{5}$ & 187\\ \hline
 II &  24.7  & 0.069 & 0.075 & $1.4\cdot10^{6}$ & 137\\ \hline
 II &  55.8  & 0.069 & 0.044 & $9.3\cdot10^{6}$ & 84\\ \hline
 II &  95.7  & 0.069 & 0.032 & $3.0\cdot10^{7}$ & 63\\ \hline

\end{tabular}
\end{center}
\end{table}

The velocity field measurements were made for steady state regimes
by a 2D particle image velocimetry (PIV) system, "Polis",
manufactured by the Institute of Thermophysics (Novosibirsk). A
CCD camera is placed above the rotating vessel in the laboratory
coordinate system as shown in Fig.\,\ref{fig_Setup_Field}. The
measurements are carried out in a confined area of the vessel, but
the rotation provided averaging for the entire horizontal
cross-section. The area of velocity measurements is shown in
Fig.\,\ref{fig_Setup_Field}. PIV measurements are done at
horizontal cross-sections for $4\leq z \leq 28$,mm in 2 mm steps
(the cylindrical coordinate system $(r,\phi,z)$ is used). We
focused on flow motions in the rotating frame, and for this
purpose the velocity field of solid-body rotation was subtracted
and all velocity fields and flow characteristics, described as
follows, are related to the motions in the rotating frame. The
time-averaged velocity fields consist of 41x30 vectors and were
averaged over 100 instant velocity fields. Then time-averaged
horizontal velocity components $v_\phi$ and $v_r$ are averaged
along the coordinate $\phi$ and the mean velocity fields at
vertical cross-section are reconstructed. The domain of PIV
measurements is restricted by the boundary layers, where the
velocity field reconstruction is hampered by optical distortions
caused by strong temperature gradients near the heaters and/or by
high velocity gradients. These boundary layers are shown in white
in figures given below. The PIV velocity measurements were
accurate to within 5\%.

\begin{figure}
\begin{center}
\includegraphics[width=.4\textwidth]{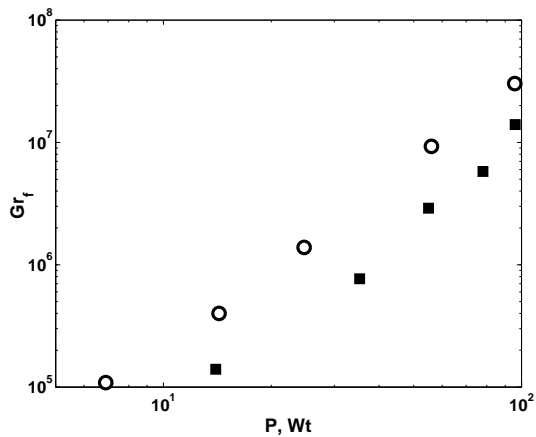}
\caption{Grasshoff number versus heating power: squares -- heater
I, circles -- heater II, $\Omega=0.069 s^{-1}$.}\label{gr_p}
\end{center}
\end{figure}

Two main factors are responsible for DR in the flows under
discussion -- convection and rotation. To characterize the
intensity of convective flows, we use the Grasshoff number ${\rm
Gr_f}$, defined in terms of layer depth $h$ and heat flux density
$q=P/S_h$ ($P$ is the power of the heater and $S_h$ is the
heater's surface area)
\begin{equation}
 {\rm {Gr_f}} = \frac{g \beta h^4 q }{ c \rho \chi \nu^2 }, \label{grass}
\end{equation}
where $g$ is the gravitational acceleration, $\beta$  is the
coefficient of thermal expansion, $c$ is the thermal capacity, and
$\rho$ is the density. Because the fluid characteristics depend on
temperature (particularly viscosity) and the mean temperature of
the layer is a function of the heating power $P$, ${\rm Gr_f}$ is
a nonlinear function of $P$. In Fig.\,\ref{gr_p} ${\rm Gr_f}$
versus $P$ is plotted on a log-log scale for both heaters. This
graph shows that the dependence is even stronger than a squared
relationship, and is similar for both heaters.

As a non-dimensional characteristic of rotation we use the Ekman
number
\begin{equation}
 E =\frac{\nu}{2 \Omega h^{2} } \label{ekman}
\end{equation}
because formation of DR strongly depends on angular momentum
exchange in viscous boundary layers. The values of parameters for
all experiments are given in Table\,\ref{tab:Exp}.

\section{Direct circulation}
\label{dir_circ}

In laboratory models of general atmospheric motions, the direct
meridional circulation is usually provided by heating the outer
(peripheral) parts of cylindrical vessels \citep{Killworth, Hide,
fultz1,fultz2}. The generated flow can be considered as a rough
model of atmospheric meridional circulation (the Hadley cell) in
which the role of latitudinal variation of $\Omega_z$ is neglected
\citep{williams2}. In our experiments, a horizontal temperature
difference is maintained by heating the periphery of the bottom
and by natural cooling at the free upper surface. We studied the
evolution of the direct meridional flow structure with variation
in the heating power and the contribution of the meridional
toroidal cell to differential rotation. All measurements are made
for a fixed rotation speed, $\Omega = 0.069 s^{-1}$.

\begin{figure}
\begin{center}
\includegraphics[width=.45\textwidth]{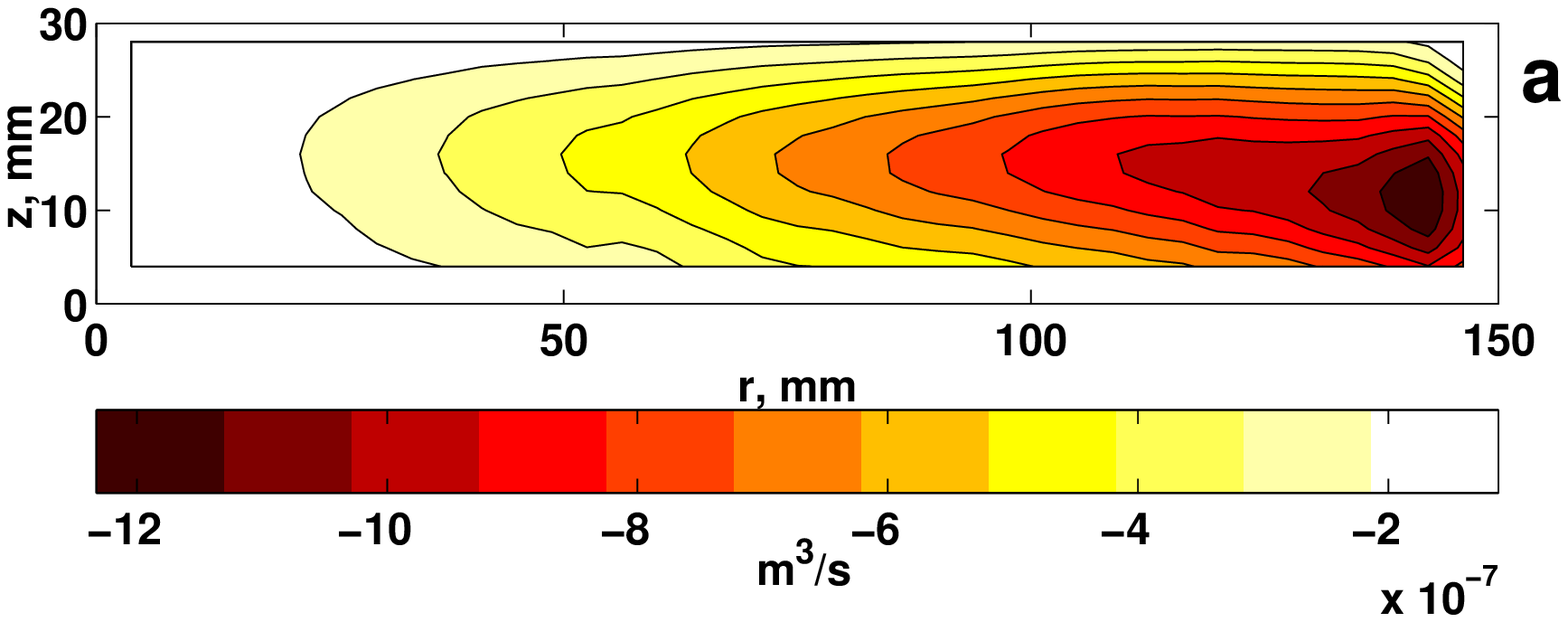}
\includegraphics[width=.45\textwidth]{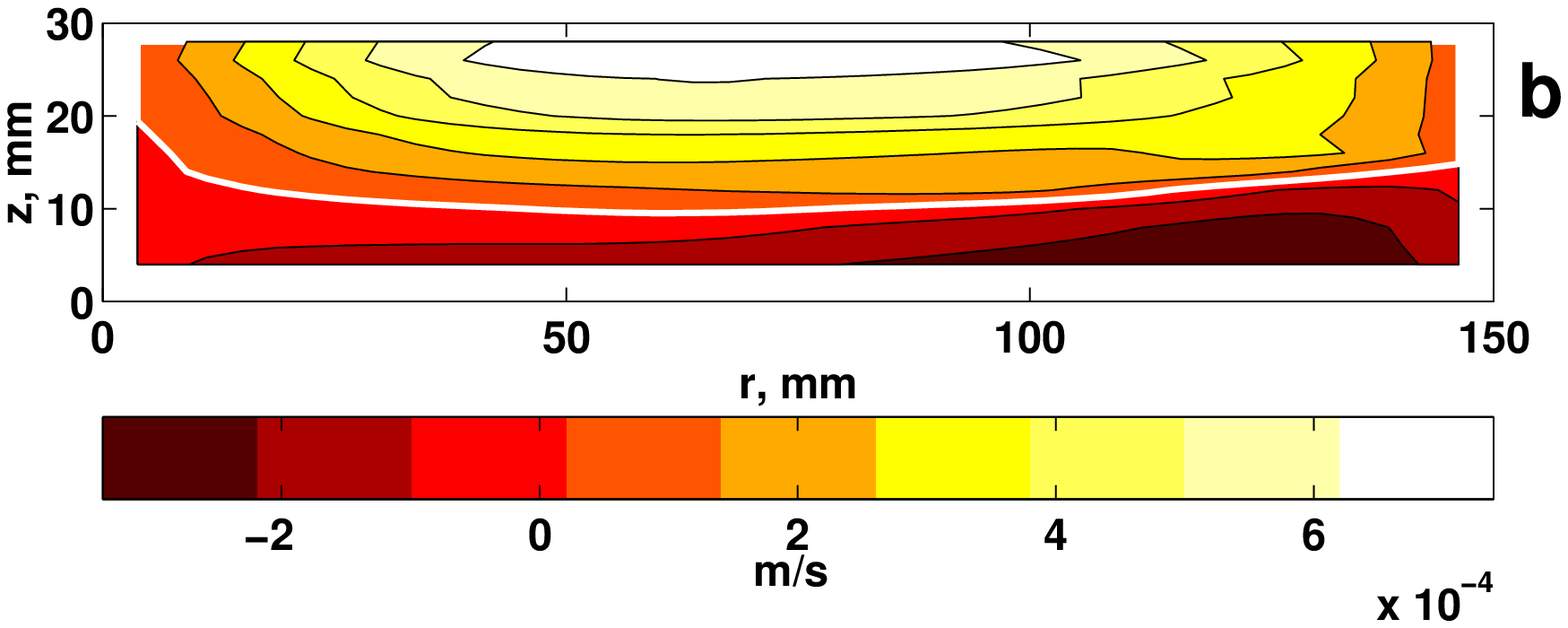}
\caption{(Color online.) Stream function of the meridional mean
velocity field (a) and mean azimuthal velocity field (b) for
direct circulation at ${\rm Gr_f}=1.4\cdot10^{5}$, $E=0.118$.
Positive values of $v_\phi$ are related to the cyclonic flow and
negative- to the anticyclonic flow. The border between the
cyclonic and anticyclonic motion is shown by the white line.}
\label{v_dir_1}
\end{center}
\end{figure}
\begin{figure}
\begin{center}
\includegraphics[width=.45\textwidth]{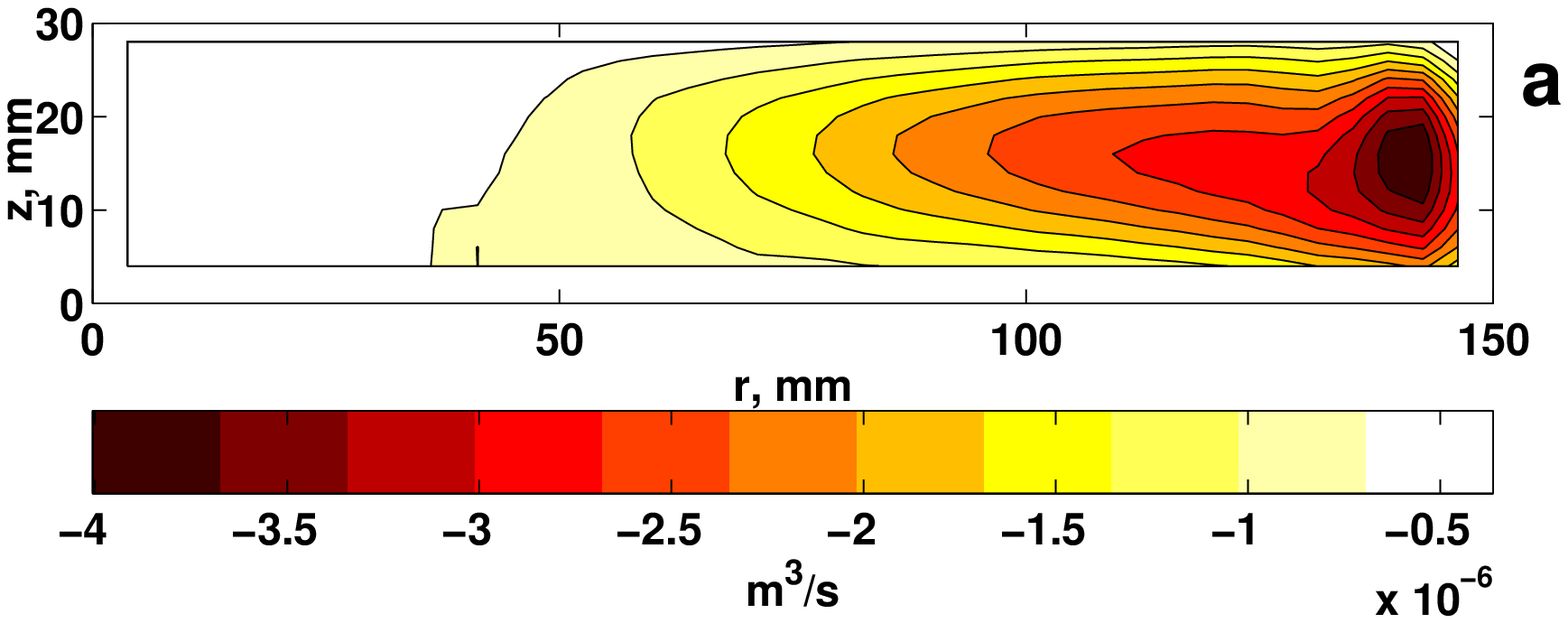}
\includegraphics[width=.45\textwidth]{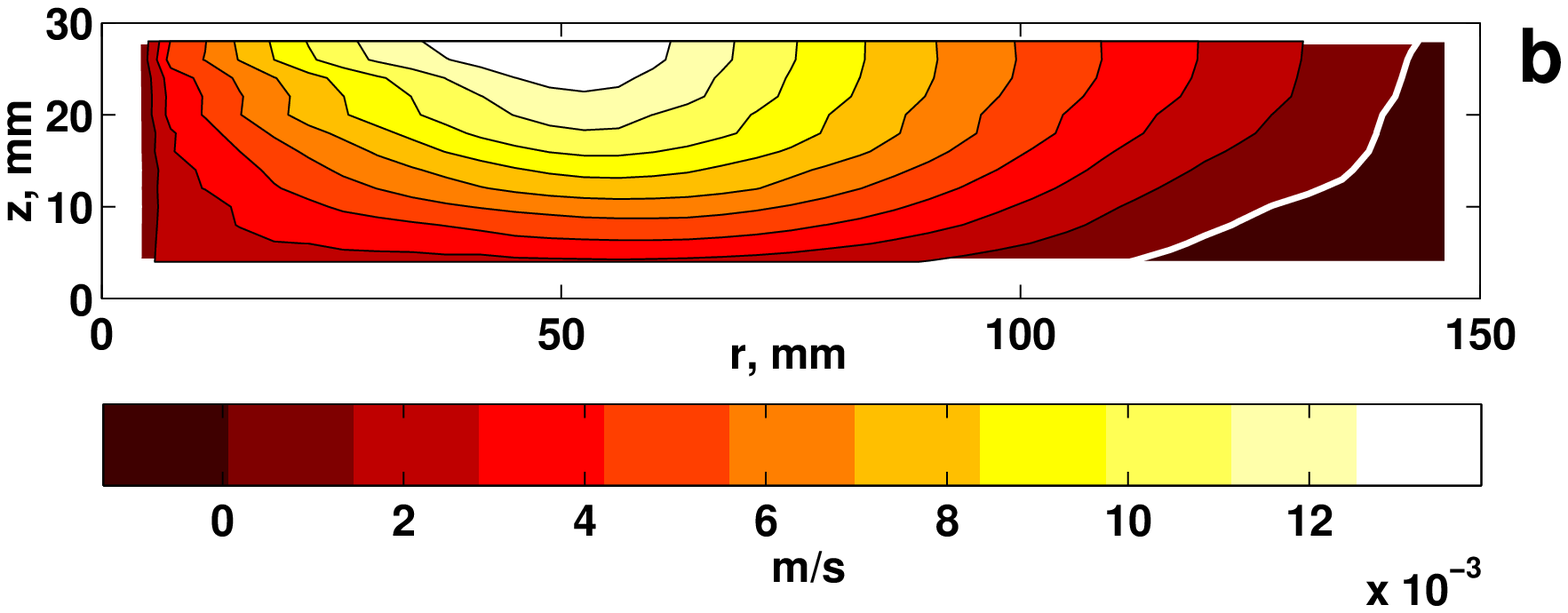}
\caption{ (Color online.) Stream function of the meridional mean
velocity field (a) and mean azimuthal velocity field (b) for
direct circulation at ${\rm Gr_f}=1.4\cdot10^{7}$, $E=0.032$. }
\label{v_dir_2}
\end{center}
\end{figure}

At weak heating the meridional flow has the structure shown in
Fig.\,\ref{circul}(a). The stream function $\psi$, defined for the
mean flow in the radial cross-section as
\begin{equation}
 \partial_z\psi =-rv_r, \qquad\qquad \partial_r\psi=rv_z, \label{psi}
\end{equation}
where $v_r$ is the radial velocity component and $v_z$ is the
vertical velocity component, is shown in Fig.\,\ref{v_dir_1},a for
${\rm Gr_f}=1.4\cdot10^{5}$, $E=0.118$. The meridional cell
occupies the whole layer providing an inward (polarward) radial
flow in the upper part and an outward flow in the lower part of
the layer. The center of the cell (the maximum of the stream
function) is localized close to the cylinder wall, where the
heater produces an intensive upward flow.

Formation of differential rotation in such a system can be
described as follows. At first, the action of the Coriolis force
on the radial flows initiates cyclonic (prograde) azimuthal flow
in the upper layer and anticyclonic (retrograde) flow near the
bottom. Then, angular momentum transport provided by meridional
circulation and angular momentum diffusion due to molecular
viscosity lead to the steady-state regime shown in
Fig.\,\ref{v_dir_1}(b). The maximum cyclonic velocity is located
near the free surface at $r \approx R/2$, and the maximum
anticyclonic velocity is located near the bottom, shifted to the
outer side.

The growth of ${\rm Gr_f}$ intensifies the meridional flow (see
Fig.\,\ref{v_dir_2}(a), where the stream function for ${\rm
Gr_f}=1.4\cdot10^{7}$ is shown). Now, the main meridional
convective cell is pressed more to the periphery of the rotating
layer. The azimuthal velocity field in Fig.\,\ref{v_dir_2}(b)
shows an absolute dominance of the cyclonic zonal circulation, and
the anticyclonic motion survives only in a small area near the
wall. The maximum of the cyclonic flow is shifted polarward
compared to its position in the previous case
(Fig.\,\ref{v_dir_1}), but the intensity of this flow grows
dramatically (the maximal azimuthal velocity grows from 0.7mm/s to
13 mm/s). The ratio of the maximum of the cyclonic flow velocity
to the velocity of the underlying solid border is more than 3
(compared to 0.1 in the previous case).

The measured velocity fields are in good qualitative agreement
with the numerical simulation of rotating convection in a wide
annular gap \citep{williams2}. The dominance of cyclonic flow in
the layer with a free surface is mainly caused by different
boundary conditions on the free surface and at the rigid bottom.
In the case with all rigid boundaries  cyclonic and anticyclonic
flow were essentially balanced \citep{williams2}.

\section{Indirect circulation}
\label{inv_circ}

In this section we describe the mean flow induced by heater II,
located at the center of the cylinder bottom. Natural convection
over the local heat sources with and without background rotation
has many geophysical and engineering applications and has been
widely studied. Most of these studies were focused on heat
transfer measurements and qualitative studies of the flow
structure. A systematic study of the dependence of the convective
flow regimes on the aspect ratio $\delta$ and the Rayleigh number
was carried out by \cite{Boubnov}. The aspect ratio was defined as
$\delta=D/h$, where $D$ is the {\it diameter of the heater}. Four
domains corresponding to different regimes were defined in the
$(\delta,{\rm Ra})$ plan: I - laminar toroidal cell; II - thermal
plumes; III - turbulent flow; and IV - transition regime.
Following this classification, all our experiments belong to case
III - turbulent flow ($\delta = 3.5$, $1.5\cdot10^{5} < {\rm Ra_h}
< 6.5\cdot10^{6}$). Here ${\rm Ra_h} = g\beta \Delta T h^3
(\nu\chi)^{-1}$.

\begin{figure}
\begin{center}
\includegraphics[width=.45\textwidth]{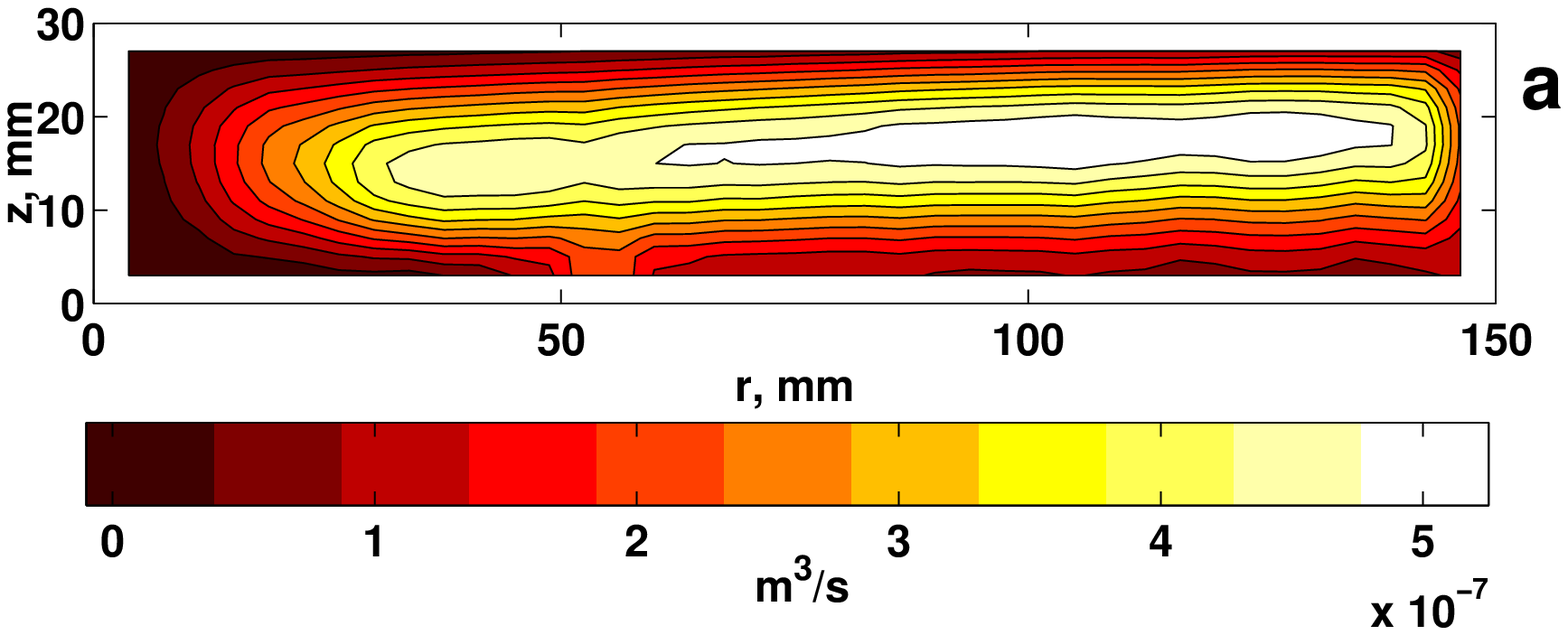}
\includegraphics[width=.45\textwidth]{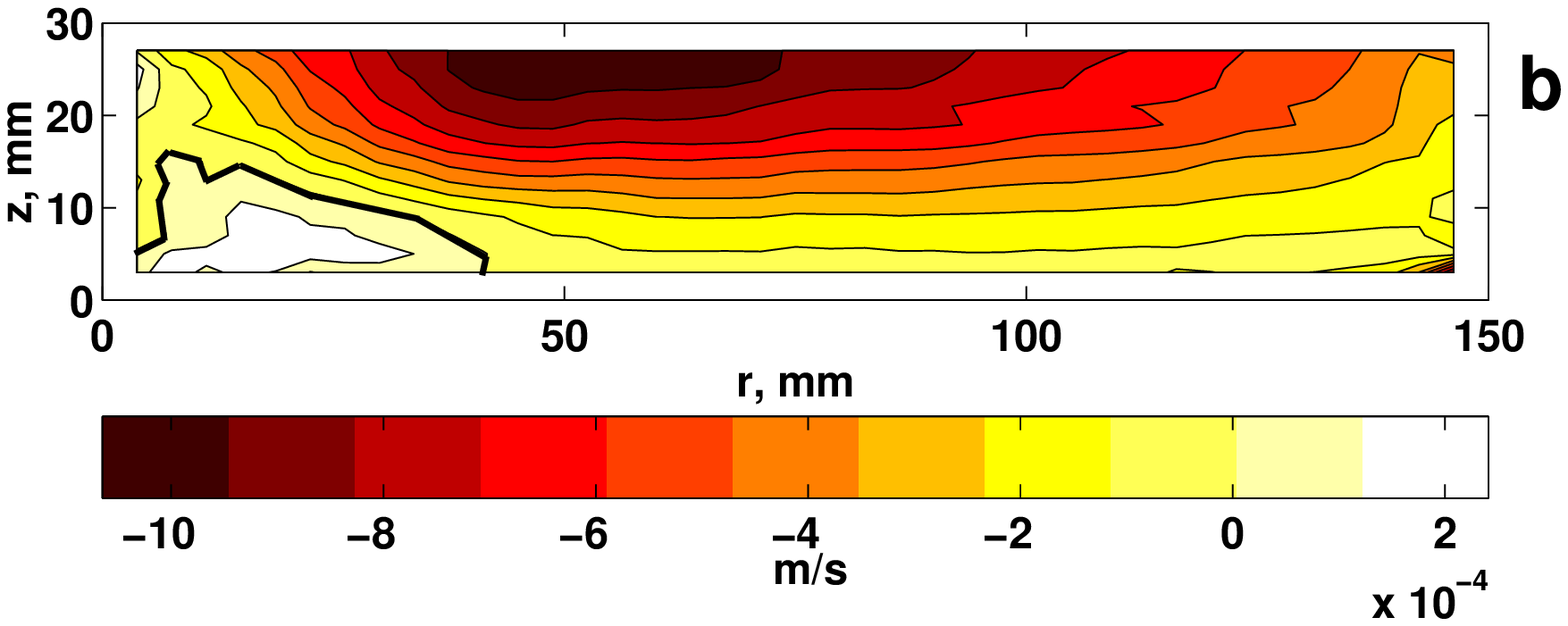}
\caption{(Color online.) Stream function of the meridional mean
velocity field (a) and mean azimuthal velocity field (b) for
indirect circulation at ${\rm Gr_f}=4.1\cdot10^{5}$, $E=0.104$.
The dashed black line marks the border between the cyclonic and
anticyclonic motion.} \label{v_inv_1}
\end{center}
\end{figure}
\begin{figure}
\begin{center}
\includegraphics[width=.45\textwidth]{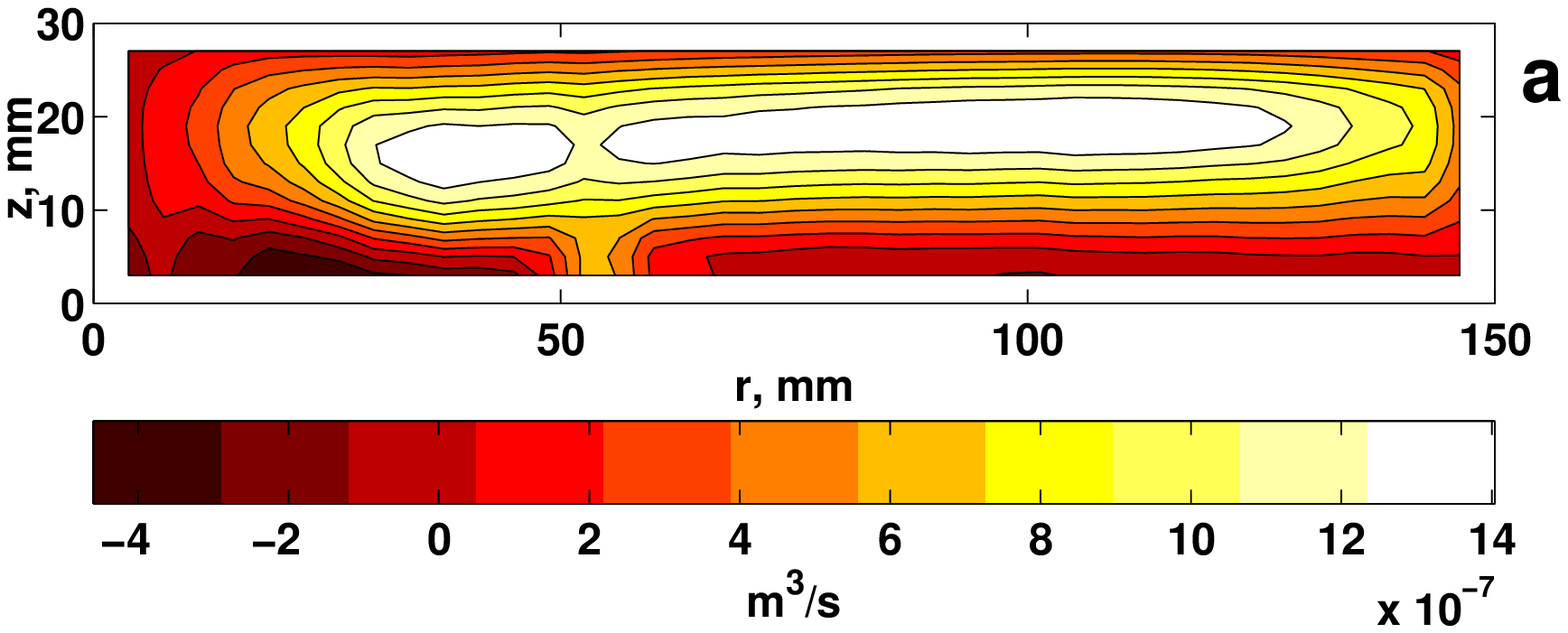}
\includegraphics[width=.45\textwidth]{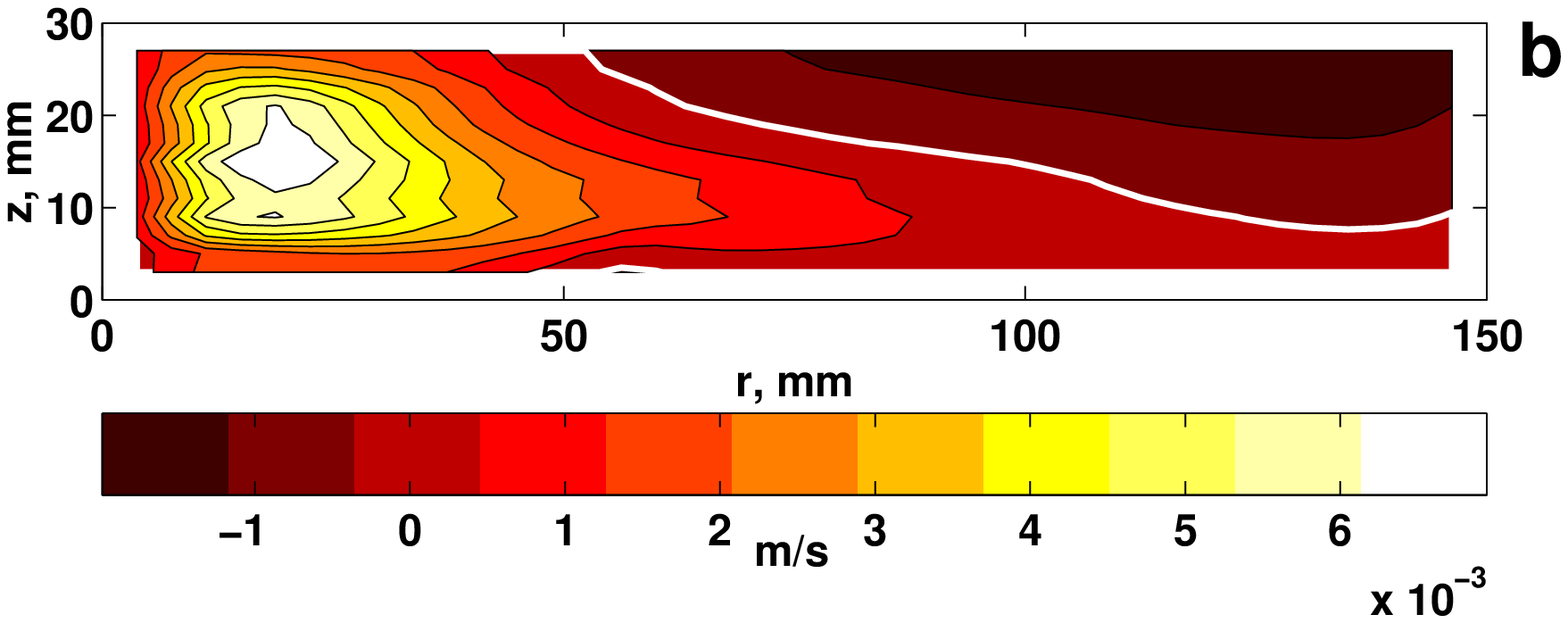}
\caption{(Color online.) Stream function of the the meridional
mean velocity field (a) and mean azimuthal velocity field (b) for
indirect circulation at ${\rm Gr_f}=3.02\cdot10^{7}$, $E=0.032$.
The solid white line marks the border between the cyclonic and
anticyclonic motion.} \label{v_inv_2}
\end{center}
\end{figure}

The structure of the mean (axisymmetric) flow is determined again
by two factors: heating power and frame rotation speed. First, we
varied the heating power at a fixed rotational speed
($\Omega=0.069$ s$^{-1}$). Local heating at the center of the
bottom generates vertical and horizontal temperature gradients.
The horizontal temperature difference provides an indirect
axisymmetric meridional cell (Fig.\,\ref{circul}(b)). The flow in
the lower part is directed inward, tending to the central bottom
region, where the heat source is located. A strong upward flux is
formed at the center above the heat source. In the upper layer,
the outward flow is directed toward the periphery. The fluid is
cooled at the free surface and finally moves downward along the
side wall. Fig.\,\ref{v_inv_1} shows the mean velocity fields at
${\rm Gr_f}=4\cdot10^{5}$. The growth of ${\rm Gr_f}$ increases
the intensity of the meridional flow, but does not appreciably
change its structure (compare panels (a) in Fig.\,\ref{v_inv_1}
and Fig.\,\ref{v_inv_2}).

Following the main direction of the meridional flow the cyclonic
motion is accumulated near the rotation axis. At low heating the
cyclonic flow occupies a small domain close to the bottom, and the
anticyclonic flows dominates in the upper layer (see
Fig.\,\ref{v_inv_1}(b)). The increase of ${\rm Gr_f}$ leads to a
substantial growth of the cyclonic velocity maximum (from 0.2 mm/s
to 7 mm/s) and weak changes of anticyclonic velocity maximum
(about 1 mm/s), compare panels (b) of
Figs.\,\ref{v_inv_1}-\ref{v_inv_2}. The cyclonic flow at high
values of ${\rm Gr_f}$ completely occupies the central part of the
layer, spreading narrow wings along the bottom up to the
periphery. The anticyclonic flow is pushed towards the side wall.

\begin{figure}
\begin{center}
\includegraphics[width=.22\textwidth]{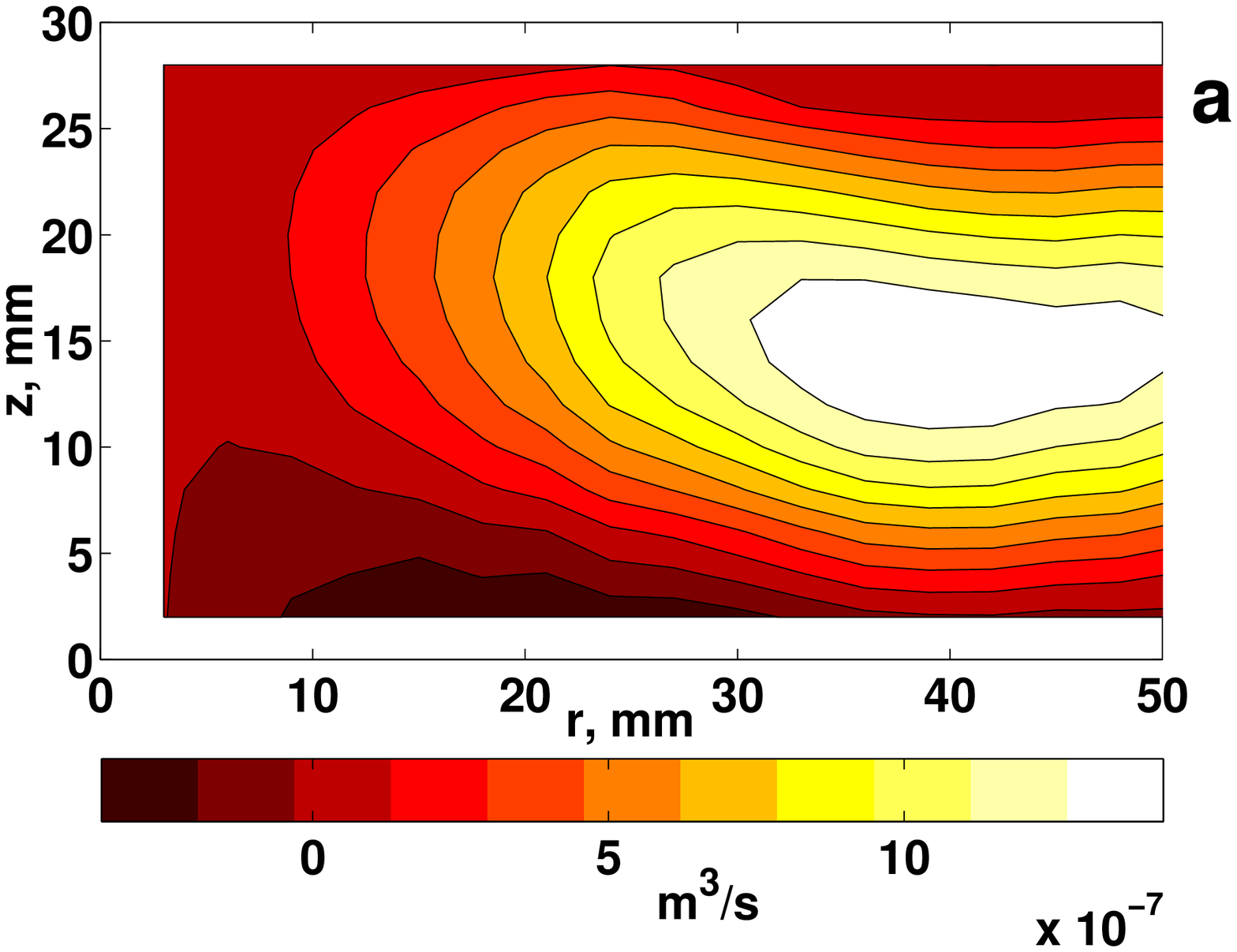}
\includegraphics[width=.22\textwidth]{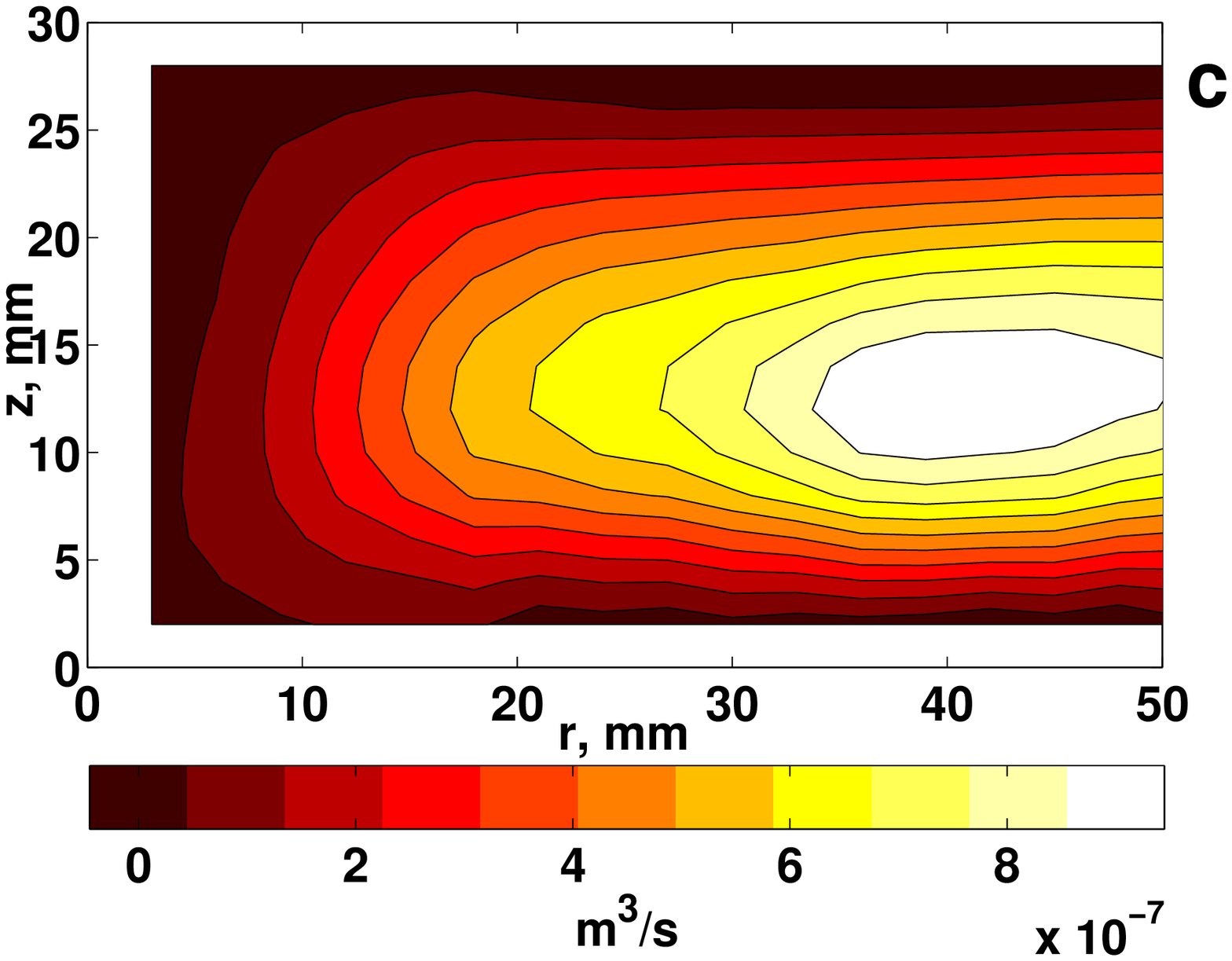}
\includegraphics[width=.22\textwidth]{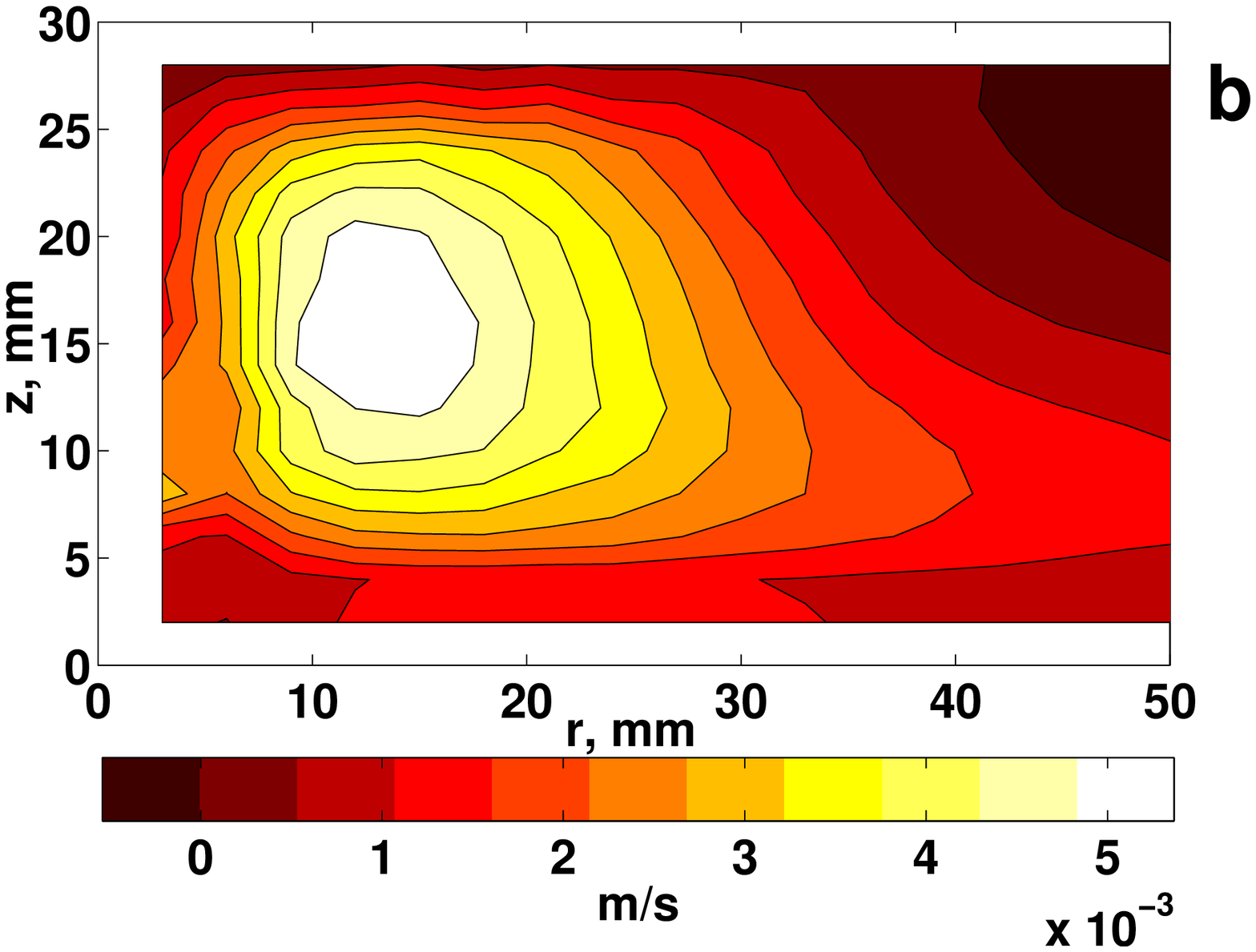}
\includegraphics[width=.22\textwidth]{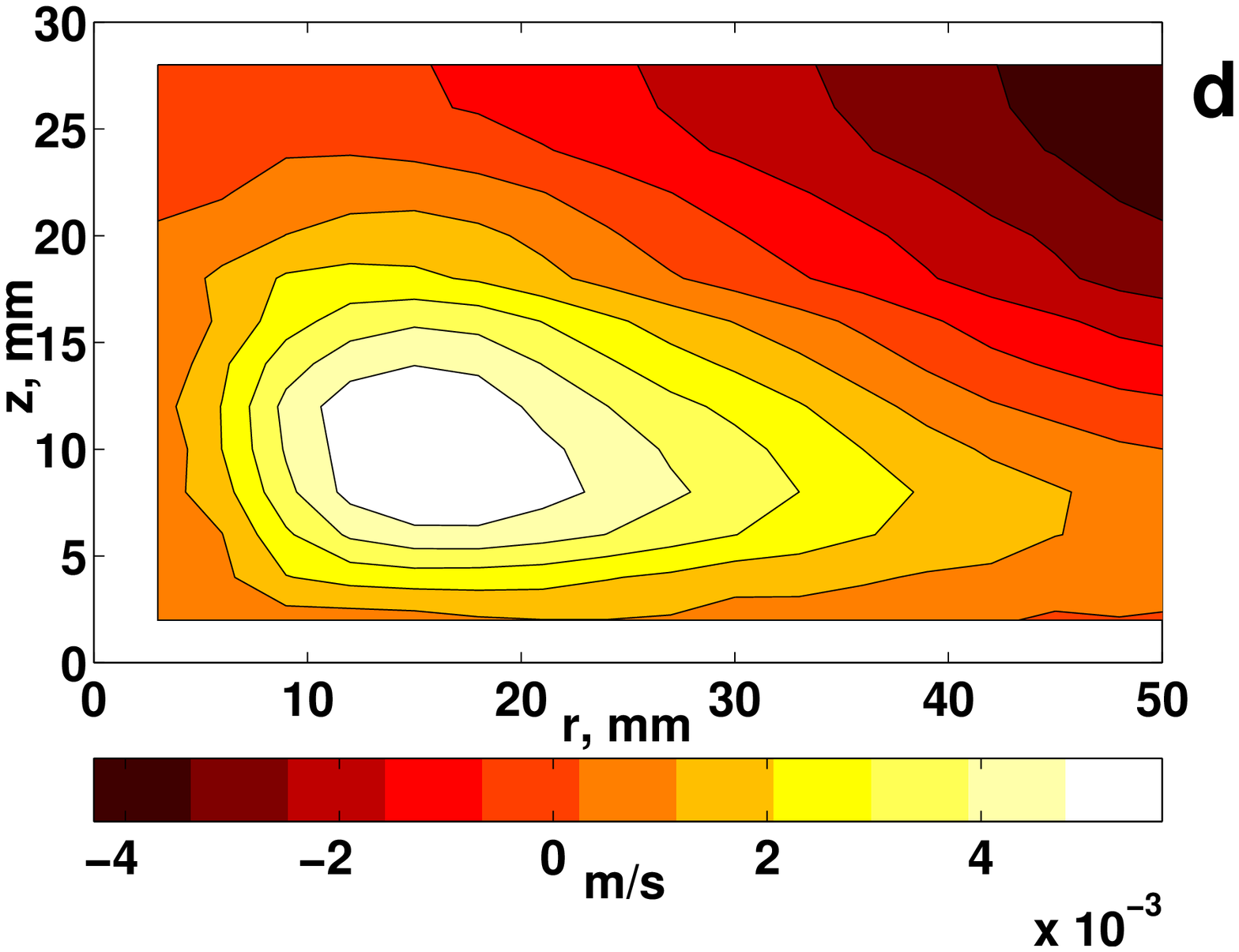}
\caption{(Color online.) Stream function of the meridional mean
velocity field (a,c) and mean azimuthal velocity field (b,d) for
indirect circulation at ${\rm Gr_f}=9.1\cdot10^{6}$ and different
rotation speed: a,b -- $\Omega=0.046$ s$^{-1}$, c,d --
$\Omega=0.161$ s$^{-1}$. Only the domain above the heater II is
shown.} \label{v_inv_3}
\end{center}
\end{figure}

The second important factor defining the flow structure is the
speed of the frame rotation. Its influence is illustrated in
Fig.\,\ref{v_inv_3}, where the velocity fields for the central
part of the layer, mostly occupied by cyclonic vortex, are shown
for ${\rm Gr_f}=9.1\cdot10^{6}$ and two different values of the
rotation speed - $\Omega=0.046$ ($E=0.066$) and $\Omega=0.161$
s$^{-1}$($E=0.019$). The growth of the frame rotation speed leads
to some decrease in intensity of the meridional circulation (see
panels (a) and (c) of Fig.\,\ref{v_inv_3}) because the increase of
the background rotation suppresses the vertical motions. The
azimuthal velocity fields for these two $\Omega$ are shown in
Fig.\,\ref{v_inv_3}(b,d). At a higher value of $\Omega$ the
cyclonic flow is confined in the lower part of the layer.

\section{Integral characteristics of differential rotation}
\label{integral}

The preceding results show that meridional circulation in both
directions leads to a pronounced DR. The direct circulation
initiates a strong cyclonic flow near the free surface at middle
radii. The anticyclonic flow in this regime is localized near the
bottom of the vessel and is weakened with increasing heating
power. The result of indirect circulation is not quite symmetric.
In this case, the cyclonic and anticyclonic flows are better
separated in radial direction -- the cyclone dominates in the
central (polar) part of the layer, spreading at high values of
heating power throughout the entire layer depth, while the
anticyclone occupies the upper part of the periphery.

\begin{figure}
\begin{center}
\includegraphics[width=.45\textwidth]{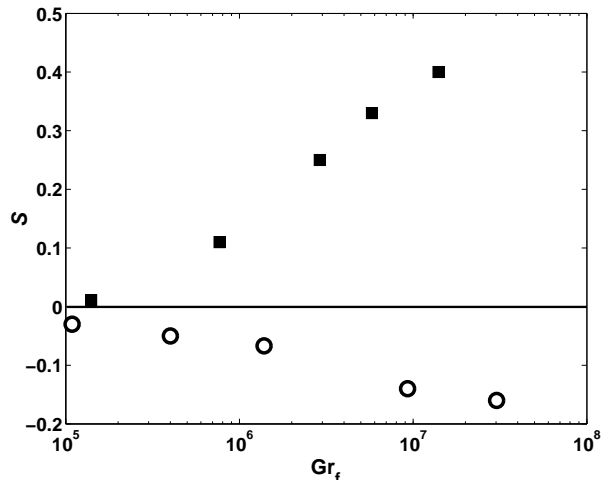}
\caption{Global super-rotation $S$ versus  ${\rm Gr_f}$ for direct
cell (squares) and indirect cell (circles) circulation,
$\Omega=0.069 s^{-1}$. } \label{Momentum}
\end{center}
\end{figure}

To quantitatively estimate the balance between the cyclonic and
anticyclonic motions, we calculated the integral angular momentum
for the whole layer; more precisely, for the whole area available
for PIV measurements. Specifically, we calculated the relative
variation $S$ of the integral momentum $L$ relative to the solid
body rotation momentum of the layer $L_s$ defined as "global
super-rotation" \citep{Read2}
\begin{equation}
S = \frac{L-L_s}{L_s}, \label{Super_global}
\end{equation}
where
\begin{eqnarray}
L = \int_{\delta_1}^{h-{\delta_2}} d z
\int_{\delta_3}^{R-{\delta_4}} r d r \int_0^{2\pi}  r v_\phi d
\phi, \\ L_s = 2\pi \int_{\delta_1}^{h-{\delta_2}} d z
\int_{\delta_3}^{R-{\delta_4}} \Omega  r^3 d r, \label{moms}
\end{eqnarray}
and $\delta_i$ are the thickness of the boundary layers which are
not accessible for PIV measurements.  The results obtained for
different values of ${\rm Gr_f}$ and a given rotation speed are
shown in Fig.\,\ref{Momentum}. The sign of the global angular
momentum is directly related to the heating mode: peripheral
heating provides the direct circulation that leads to the growth
of the integral angular momentum of the fluid layer. On average,
the fluid layer rotates faster than the background; therefore,
this is a case of super-rotation. The maximum value of $S$
achieved in the experiment is about 0.4. In contrast, central
heating provides the indirect circulation that reduces the
integral angular momentum; on average the layer rotates more
slowly than the background (sub-rotation). Note that the increase
in angular momentum at direct circulation is much greater than its
decrease at indirect circulation for the same value of the heating
power. The lowest value of $S$ in the case of the indirect cell
was $S\approx -0.16$.

\begin{figure}
\begin{center}
\includegraphics[width=.45\textwidth]{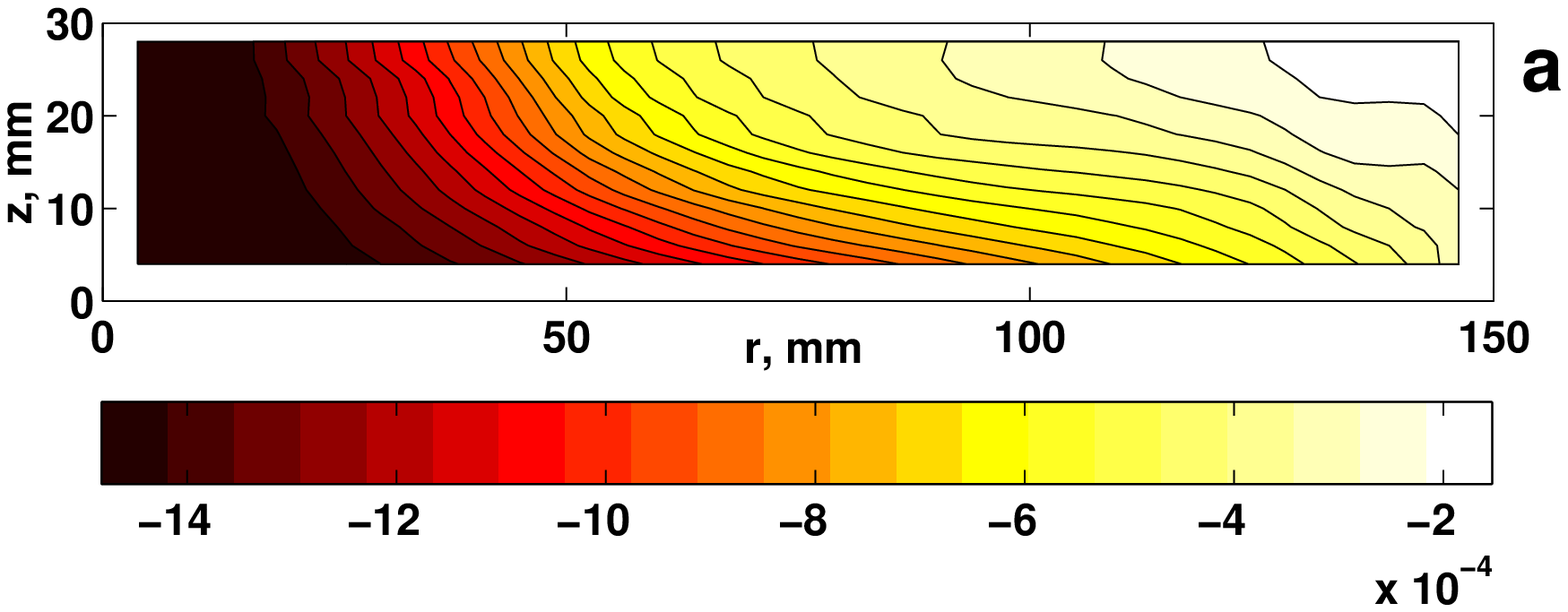}
\includegraphics[width=.45\textwidth]{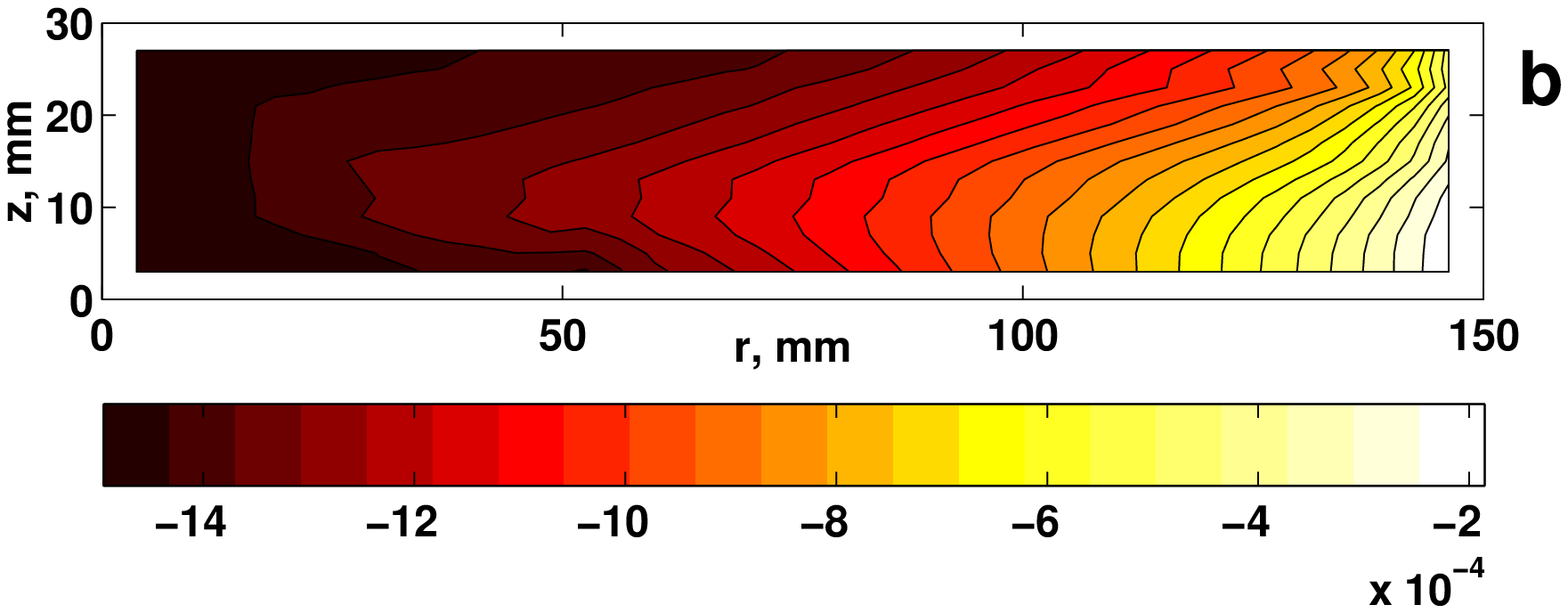}
\caption{(Color online.) Local super-rotation $s$: (a)-- direct
circulation for ${\rm Gr_f}=1.4\cdot10^{7}$, $E=0.032$; (b)--
indirect circulation for ${\rm Gr_f}=3.02\cdot10^{7}$, $E=0.032$.
} \label{local_s_field}
 \end{center}
 \end{figure}

Useful information might give the so-called "local super-rotation"
$s$ \citep{Read2}, defined as
\begin{equation}
s = \frac{v_\phi r}{ \Omega R^2}-1. \label{local}
\end{equation}
It shows the excess of angular momentum in comparison with highest
possible angular momentum of the fluid element in a solid-body
rotation state. The distribution of $s$ for different types of
meridional circulation is shown in Fig.\,\ref{local_s_field}. This
figure shows that even for high values of Grasshoff number  the
local super-rotation $s$ is negative everywhere (sub-rotation).
Note, that numerical simulation in case of stress-free sidewalls
showed that there is noticeable area with $s>0$ \citep{Read2}.
This proves the strong influence of the side wall boundary
conditions on the distribution of angular momentum.

Angular momentum redistribution in the fluid layer occurs due to
the action of net torque due to friction at the bottom and side
walls (the upper surface can be considered stress-free, and the
torque on the upper surface is negligible). In the case of direct
meridional circulation the anticyclonic flow in the lower layer
generates torque that injects angular momentum into the layer.
Meridional flow provides the transfer of angular momentum from the
lower to upper part of the layer. The sink of the angular momentum
occurs in the viscous boundary layers occupied by the cyclonic
flow. In the case of an indirect cell the sink of the angular
momentum is located in the central part of the bottom occupied by
an intensive cyclonic vortex, and the source of the angular
momentum is located at the upper part of the side wall - the area
of anticyclonic flow.

Thus, the flow structure and the imbalance between cyclonic and
anticyclonic motion observed in the experiment is a consequence of
different boundary conditions: non-slip on rigid boundaries, and
stress-free on the surface. This agrees with general conclusions
made from numerical simulations \citep{williams2} and
\citep{Read2}.

Note that some discrepances exist between our experimental results
and the analysis of the azimuthal circulation made in
\citep{Read2}, where the flow structure analysis was made through
dimensionless parameter $Q$ which was defined using the aspect
ratio, Ekman number and Rayleigh number Ra,
\begin{equation}
Q = {\rm {Ra}}^{-1/2}E^{-1}\varepsilon^{-3/2}, \label{Q}
\end{equation}
\begin{eqnarray}
{\rm {Ra}} = \frac{g \beta L^3 \Delta T }{ \chi \nu } , \label{Ra}
\end{eqnarray}
where $L$ is the horizontal length scale and $\Delta T $ is the
horizontal temperature drop. $Q$ describes the ratio between
thickness of thermal boundary layer (non-rotating) and that of the
Ekman boundary layer. It was shown that global super-rotation
depends on $Q$ in the follow way: the super-rotation is almost
constant for slow rotation ($Q<(\varepsilon/{\rm Pr})^2$); it
increases at moderate rotation ($(\varepsilon/{\rm Pr})^2Q<1$),
and it decreases quickly at rapid rotation ($Q>1$). Calculating
$Q$ for all measured experimental regimes (taking $L=R$), we found
that $Q$ changes very weakly (see Table\,\ref{tab:Gr_Q}). Thus,
the whole variety of observed regimes exist under similar values
of $Q$ and belong to the case of "moderate" rotation.

The independence of $Q$ on ${\rm Gr_f}$ can be easily understood
if we expand the expression for $Q$ and rewrite it in the
following form
\begin{equation}
Q = 2 \Omega \left( \frac{h}{g \beta \Delta T}\cdot \frac{1}{{\rm
Pr}} \right)^{1/2}. \label{Q_expand}
\end{equation}
In our experiments, all parameters from (\ref{Q_expand}) except
${\rm Pr}$ and $\Delta T$ were constant. The growth of ${\rm
Gr_f}$ leads to an increase of the mean temperature of the fluid
$T$ and $\Delta T$, but to the decrease in the value of ${\rm
Pr}$. The decrease of ${\rm Pr}$ with growth of $T$ is caused by
strong temperature dependence of oil viscosity $\nu$. As a result
$({{\rm Pr} \Delta T})^{-1/2}$ and parameter $Q$ are almost
independent of ${\rm Gr_f}$. Note that the simulations of
\cite{Read2} were performed for $\varepsilon=1$, ${\rm Pr}=10$, a
fixed temperature drop (heating) and varying Ekman number. In our
experiments, the aspect ratio was smaller ($\varepsilon=0.2$), the
Prandtl number an order of magnitude larger, the rotation speed
fixed. The heating power was the main governing parameter, which
provided the variety of regimes.

\begin{table}
\begin{center}
\caption {Grasshoff number and the parameter $Q$ for all
experiments} \label{tab:Gr_Q}

\begin{tabular}{|c|c|c|c|c|c|} \hline
  ${\rm Gr_f}$ & $1.4\cdot10^{5}$ & $7.7\cdot10^{5}$ & $2.9\cdot10^{6}$ & $5.8\cdot10^{6}$ & $1.4\cdot10^{7}$ \\ \hline
  $Q$ & 0.0082 & 0.0066 & 0.0066 & 0.0063 & 0.007 \\ \hline
\end{tabular}
 \end{center}
 \end{table}

\begin{figure}
\begin{center}
\includegraphics[width=.4\textwidth]{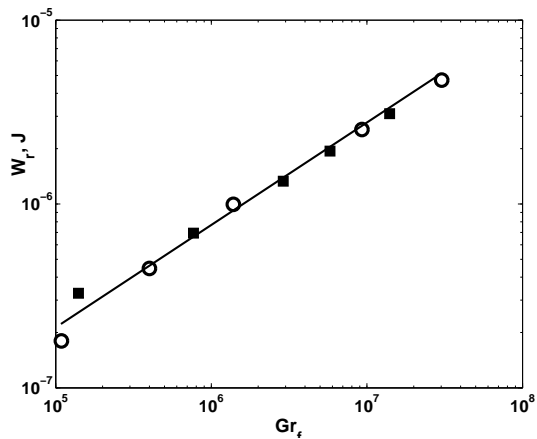}
\caption{Total energy of the meridional flow versus Grasshoff
number for direct (squares) and indirect (circles) circulation,
$\Omega=0.069 s^{-1}$. The solid line shows the slope for ${\rm
W_r}\sim \sqrt {\rm Gr}$.
 } \label{en_merid}
 \end{center}
 \end{figure}

As previously stated, the structure of the meridional flow is
different at direct and indirect circulations (compare
Fig.\,\ref{v_dir_2},a and Fig.\,\ref{v_inv_2},a). To estimate the
energy balance, we compared the total energy of the meridional
flow ${\rm W_r}$ for different kinds of circulation. Our
measurements showed that the energy of the indirect meridional
flow (provided by heater II) is about two times higher than the
energy of the direct circulation (provided by heater I) for the
same heating power. Therefore, for comparison of the energy
budgets of the direct and indirect cells we used the Grasshoff
number ${\rm Gr_f}$ and plotted the energy of the meridional
circulation versus ${\rm Gr_f}$. Fig.\,\ref{en_merid} shows that
despite different structures of meridional circulation for the
direct and indirect cells ${\rm W_r}$ for both cells is well
correlated with the Grasshoff number. The function ${\rm
W_r(Gr_f})$ is given in log-log scale and displays a power law
close to "1/2".

\begin{figure}
\begin{center}
\includegraphics[width=.4\textwidth]{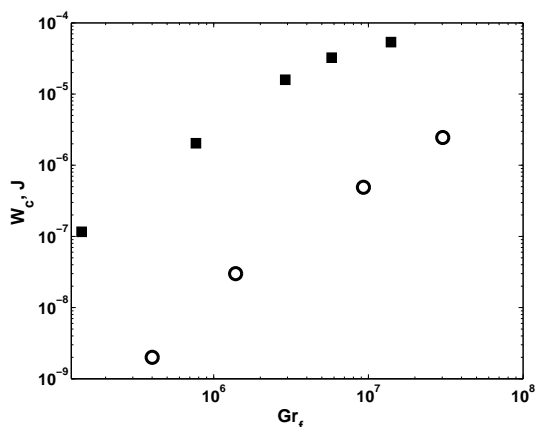}
\caption{Total energy of the cyclonic flow at direct (squares) and
indirect (circles) circulation versus the Grasshoff number,
$\Omega=0.069 s^{-1}$. } \label{ecgr}
 \end{center}
 \end{figure}

We then analyzed how the energy of the cyclonic and anticyclonic
motions ${\rm W_c}$ and ${\rm W_a}$ depend on the type of
circulation and on the Grasshof number. Direct circulation results
in a much more intensive cyclonic flow -- than indirect
circulation at the same ${\rm Gr_f}$. However, in spite of the
great difference in the structures and intensities of the cyclonic
flow in the direct and indirect cells the dependence of ${\rm
W_c}$ on the Grasshoff number is similar in both cases (see
Fig.\,\ref{ecgr}, where the energy of the cyclonic motion is shown
versus the Grasshoff number for both circulations).

\begin{figure}
\begin{center}
\includegraphics[width=.4\textwidth]{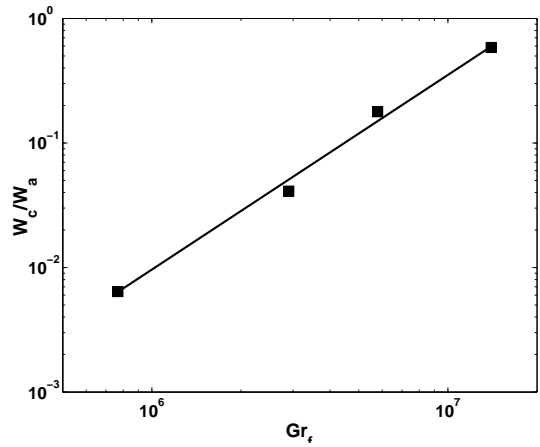}
\caption{Ratio of the cyclonic motion energy  to the energy of
anticyclonic motion under indirect circulation, $\Omega=0.069
s^{-1}$. } \label{eaec}
 \end{center}
 \end{figure}

We could not obtain a similar picture for anticyclonic flow,
because in the case of direct circulation the anticiclonic motion
is mainly located in the boundary layer and could not be
reconstructed with sufficient accuracy. Fig.\,\ref{eaec} shows the
ratio of the cyclonic to anticyclonic flow energy for the indirect
circulation. From this graph it is evident that in the considered
range of parameters, this ratio grows linearly with the Grasshoff
number. Thus, if this tendency remains valid at higher values of
${\rm Gr}_f$, the energy of the cyclonic flows will surpass the
energy of the antiyclonic flow at ${\rm Gr}_f > 2\cdot 10^7$.

Finally, we considered the direct characteristics of differential
rotation. Note that DR is one of the major mechanisms of magnetic
field generation in cosmic bodies; namely, DR is responsible for
generation of the toroidal (azimuthal) magnetic field from the
poloidal field. The efficiency of this mechanism depends on the
gradient of the azimuthal velocity. The nondimensional
characteristic of a dynamo process based on DR -- the so-called
"dynamo number" includes some characteristic value of the gradient
of the mean azimuthal velocity (in the radial or meridional
direction) in the domain where the dynamo action is localized
\citep{zrs83}. Since we studied the DR outside a special dynamo
problem, we use, as characteristics of DR, the mean values of the
gradients $\partial_r v_\phi$ and $\partial_z v_\phi$ calculated
over the domain of reconstruction.
\begin{eqnarray}
D_r = 2\pi\int_{\delta_1}^{h-{\delta_2}} d z
\int_{\delta_3}^{R-{\delta_4}} r  (  \partial_r v_\phi)d r ,
\\ D_z = 2\pi \int_{\delta_1}^{h-{\delta_2}} d z
\int_{\delta_3}^{R-{\delta_4}} r  ( \partial_z v_\phi)d r.
\label{D}
\end{eqnarray}

\begin{figure}
\begin{center}
\includegraphics[width=.4\textwidth]{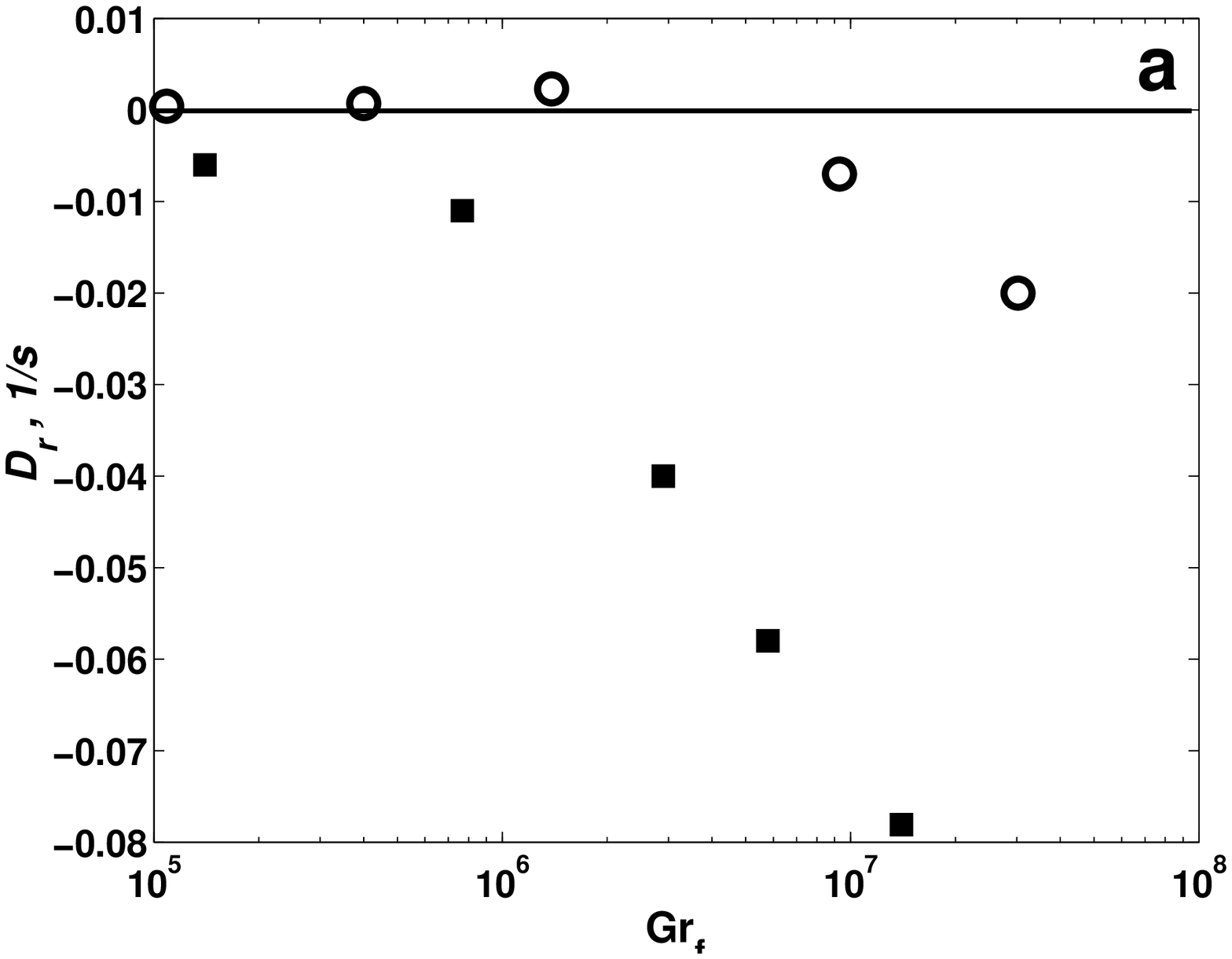}
\includegraphics[width=.4\textwidth]{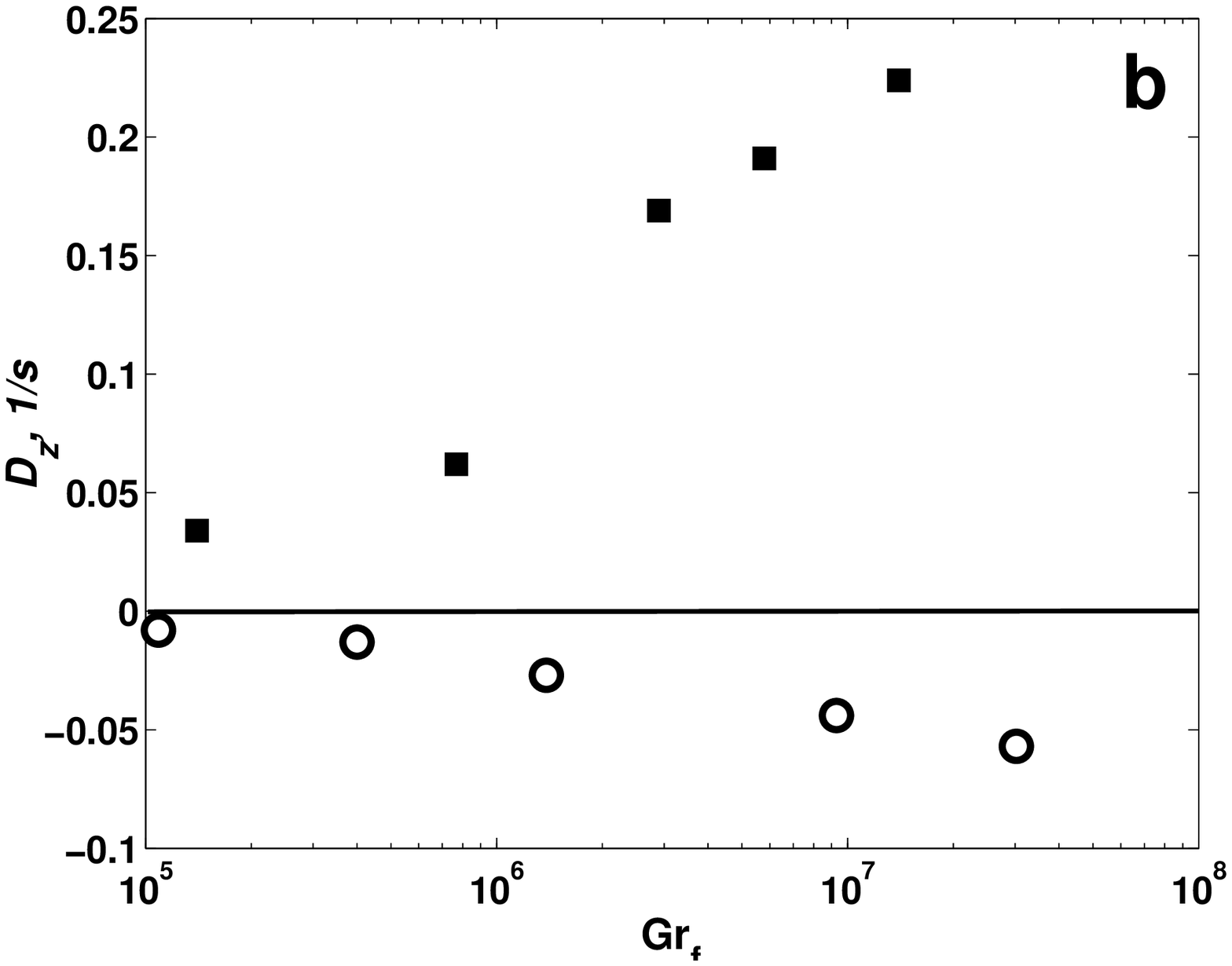}
\caption{Mean radial $D_r$ (a) and vertical $D_z$ (b) gradients of
azimuthal velocity  for direct (squares) and indirect (circles)
circulation versus the Grasshoff number for, $\Omega=0.069
s^{-1}$. } \label{difrot}
 \end{center}
\end{figure}

The results are shown in Fig.\,\ref{difrot} as functions of the
Grasshoff number for both circulations. Panel (a) of this figure
shows that the mean radial gradient is negative for direct
circulation (the central part rotates faster than the periphery),
and that its value grows with the Grasshoff number. For indirect
circulation, the mean radial gradient is close to zero at low
Grasshoff number (the cyclonic and anticyclonic motions are
separated vertically, but not radially) and becomes negative at
high ${\rm Gr_f}$. Then, the radial DR is weaker as in case of
direct circulation, but the central part also rotates faster than
the periphery. The behavior of the mean vertical gradient is quite
different for two circulations: a positive vertical gradient
increases with a growing Grasshoff number (approximately as
$\log(\rm Gr_f)$) at direct circulation, whereas a weak negative
gradient characterizes the indirect circulation.

\section{Summary}
\label{diss}

We studied the formation of DR for direct and indirect meridional
cells. In the case of a direct cell, the cyclonic flow prevails
and spans nearly the entire layer except for a relatively small
zone at the periphery near the bottom. The indirect cell is
characterized by competition between the cyclonic flow in the
central part and anticyclonic flow at the periphery. The ratio
$W_c/W_a$ grows with ${\rm Gr_f}$ and approaches unity at ${\rm
Gr_f} \approx 2\cdot 10^7$.

The structure of azimuthal flows and their intensities are
determined by the difference between friction force torques at the
rigid boundaries and the free surface. The condition of a
steady-state regime is zero net torque. Torque due to frictional
forces at the solid boundaries serves as a source of the angular
momentum in the anticyclonic flow area and as a sink in the
cyclonic flow area. During transition to a steady-state regime the
net torque is nonzero and leads to variation in overall angular
momentum. Direct circulation leads to the growth of overall
angular momentum, whereas indirect circulation causes it to
decrease. At the same ${\rm Gr_f}$ the increase in angular
momentum at direct circulation is much stronger than its decrease
at indirect circulation.

DR is caused by the meridional transport of angular momentum. We
have shown that the energy of the meridional circulation grows
with the Grasshoff number as $\sqrt{\rm Gr_f}$ at both directions
of circulation. We characterize the differential rotation by the
mean values of radial and vertical gradients of the azimuthal
velocity. Direct circulation provides a strong negative mean
radial gradient, which means that the central part rotates faster.
The same tendency exists at indirect circulation, but the gradient
is much smaller or even close to zero at low Grasshoff numbers.
Direct circulation gives rise to a pronounced vertical gradient
(positive), which grows with Grasshoff number. Indirect
circulation provides a weak negative gradient. This difference
follows from the very structure of the flow: direct circulation
provides a large cyclonic area localized above the anticyclonic
flow, and indirect circulation leads to substantial separation of
these two area in the radial direction (the cyclonic flow is
localized close to the rotation axis and the anticyclonic flow is
concentrated near the side wall).

\section*{Acknowledgment}

Financial support from RFBR under grants 07-01-92160 and
07-05-00060 is gratefully acknowledged.

\end{document}